\documentclass[showpacs,reprint]{revtex4}
\usepackage{amsfonts}
\usepackage{amssymb}
\usepackage{amsmath}
\usepackage{hyperref}
\usepackage{latexsym}
\usepackage[dvips]{graphicx}
\usepackage{epsf}
\usepackage{multirow}

\begin{document}

\title{On possible origin of an anisotropy in the speed of light in vacuum}
\author{ E.P. Savelova}
\date{}

\begin{abstract}
We assume the spacetime foam picture in which vacuum is filled with virtual
wormholes. In the presence of an external field the distribution of
wormholes changes. We consider an anisotropic distribution of wormholes and
analyze its relation to the speed of light. We show that speed of light
acquires an anisotropic character and save the normal dispersion a gas of
virtual wormholes may possess also an anomalous dispersion, i.e., when the
light velocity exceeds that in the vacuum.
\end{abstract}

\maketitle

\address{Dubna International University of Nature, Society and Man }

\section{Introduction}

An actual wormhole requires the violation of averaged null energy condition
(ANEC) \cite{VisX} and is commonly supposed to be forbidden in classical
physics. However, the situation changes when we consider a virtual wormhole.
It represents a quantum topology fluctuation (tunnelling between different
topologies) which takes place at very small (Planckian) scales and lasts for
a very short period of time (the so-called spacetime foam, e.g., see \cite%
{wheelerX,H78X}). It does not obey to the Einstein equations and, therefore,
ANEC cannot forbid the origin of such an object. In other words, virtual
wormholes violate readily ANEC.

Virtual wormholes may be described by Euclidean wormhole configurations
which were first suggested in Refs. \cite{Loss1} (see also ref. \cite{Loss2}%
) where it was proposed that they can lead to loss of quantum coherence. It
was latter shown though \cite{Lambda} that the effects of such wormholes can
be absorbed into a redefinition of coupling constants of the low energy
theory and, therefore, quantum coherence is not lost. However, as it was
recently shown \cite{KS10X,S13} virtual wormholes may play an important role
in particle physics at very high energies, since they may introduce in a
natural way the cutoff at very small scales and may remove divergencies in
quantum field theory. \ Moreover, the existence of virtual wormholes may be
important for the possibility to form wormhole-like objects at laboratory
scales \cite{KS13a}. Indeed, a coherent set of virtual wormholes may work as
an actual wormhole \cite{S13} (see also ref. \cite{Knot}). In particular, by
applying an external classical field one may govern the intensity of such
fluctuations and try to organize an artificial wormhole \cite{KS13a}.

In the present paper we consider the situation when additional wormholes are
incoherent but simply have an anisotropic distribution that leads to an
anisotropy in the propagation of light in vacuum and analyze the resulting
dispersion relations. We explicitly demonstrate that virtual wormholes may
possess both, the normal and anomalous dispersion, i.e., when the resulting
speed of light is smaller or bigger than the speed of light in vacuum.

\section{Virtual wormhole}

In what follows we shall use some of results of Ref. \cite{KS10X}.
A virtual wormhole can be described by the Euclidean wormhole
configuration as follows (for the Euclidean approach see also
Refs. \cite{H78X,bookX}). The simplest virtual wormhole may be
constructed by the gluing procedure. Consider the metric ($\alpha =1,2,3,4$)%
\begin{equation}
ds^{2}=h^{2}\left( r\right) \delta _{\alpha \beta }dx^{\alpha }dx^{\beta },
\label{X_wmetr}
\end{equation}%
where
\begin{equation}
h\left( r\right) =1+\theta \left( a-r\right) \left( \frac{a^{2}}{r^{2}}%
-1\right)
\end{equation}%
and $\theta \left( x\right) $ is the step function. Such a wormhole has
vanishing throat length. Indeed, in the region $r>a$, $h=1$ and the metric
is flat, while the region $r<a$, with the obvious transformation $y^{\alpha
}=\frac{a^{2}}{r^{2}}x^{\alpha }$, is also flat for $y>a$. Therefore, the
regions $r>a$ and $r<a$ represent two Euclidean spaces glued at the surface
of a sphere $S^{3}$ with the center at the origin $r=0$ and radius $r=a$.
Such a space can be described with the ordinary double-valued flat metric in
the region $r_{\pm }>a$ by
\begin{equation}
ds^{2}=\delta _{\alpha \beta }dx_{\pm }^{\alpha }dx_{\pm }^{\beta },
\label{X_wmetr2}
\end{equation}%
where the coordinates $x_{\pm }^{\alpha }$ describe two different sheets of
space. We point out that in the quasi-classical region a virtual wormhole
may be taken as a solution of the Euclidean Einstein equations and the
function $h$ should be smooth. In particular, the choice $%
h(r)=(r^{2}+a^{2})/r^{2}$ corresponds to the so-called Bronikobv-Ellis type
metric
\begin{equation*}
ds^{2}=dR^{2}+(R^{2}+4a^{2})d\Omega ^{2},
\end{equation*}%
where $d\Omega ^{2}$ is the angle part of the metric, $R=r-a^{2}/r$ and $%
-\infty <R<\infty $.

Let identify the inner and outer regions of the sphere $S^{3}$ and construct
a wormhole which connects regions in the same space (instead of two
independent spaces). This is achieved by gluing the two spaces in (\ref%
{X_wmetr2}) by motions of the Euclidean space. If $R_{\pm }$ is the position
of the sphere in coordinates $x_{\pm }^{\mu }$, then the gluing is the rule%
\begin{equation}
x_{+}^{\mu }=R_{+}^{\mu }+\Lambda _{\nu }^{\mu }\left( x_{-}^{\nu
}-R_{-}^{\nu }\right) ,  \label{X_gl}
\end{equation}%
where $\Lambda _{\nu }^{\mu }\in O(4)$, which represents the composition of
a translation and a rotation of the Euclidean space. In terms of common
coordinates such a wormhole represents the standard flat space in which the
two spheres $S_{\pm }^{3}$ (with centers at positions $R_{\pm }$) are glued
by the rule (\ref{X_gl}). We point out that the physical region is the outer
region of the two spheres. Thus, in general, the wormhole is described by a
set of parameters: the throat radius $a$, positions of throats $R_{\pm }$ ,
and rotation matrix $\Lambda _{\nu }^{\mu }\in O(4)$.

\section{Green function for a single wormhole connecting two spaces}

In the Euclidean field theory the Green function for a scalar field obeys
the Laplace equation
\begin{equation}
\left( -\Delta +m^{2}\right) G(x,x^{\prime })=\delta (x-x^{\prime })
\label{weq}
\end{equation}%
where $m$ is the mass, $\Delta =\frac{1}{\sqrt{g}}\partial _{\alpha }(\sqrt{g%
}g^{\alpha \beta }\partial _{\beta })$, and $g_{\alpha \beta }$ is the
metric. In the Euclidean space the metric is flat $g_{\alpha \beta }=\delta
_{\alpha \beta }$ and the above equation has the well-known solution $%
G_{0}(x,x^{\prime })=\frac{m^{2}}{4\pi ^{2}}\frac{K_{1}(mr)}{mr}$, where $%
r^{2}=(x-x^{\prime })^{2}$ and $K_{1}(x)$ is the modified Bessel function.

When considering a space with a wormhole the metric cannot be chosen
everywhere flat and the exact form of the Green function depends on the
specific structure of the wormhole. In the present section we construct the
Green function for the simplest wormhole connecting two Euclidean spaces
which is described by the $O(4)$ invariant metric (\ref{X_wmetr}). In this
case the equation (\ref{weq}) admits the exact solution, e.g., see details
in Ref. \cite{S12ef}.

Indeed, consider four-dimensional spherical coordinates $r$, $\chi $, $%
\theta $, $\phi $ in which the square of the element of length is%
\begin{equation*}
ds^{2}=h^{2}\left( r\right) \left( dr^{2}+r^{2}\left[ d\chi ^{2}+\sin
^{2}\chi \left( d\theta ^{2}+\sin ^{2}\theta d\phi ^{2}\right) \right]
\right) .
\end{equation*}%
Then the angular part of the Green function can be decomposed in terms of
four-dimensional spherical harmonics as%
\begin{equation*}
G(x,x^{\prime })=\sum_{n=1}^{\infty
}\sum_{l=0}^{n-1}\sum_{m=-l}^{l}Q_{nlm}^{\ast }(\Omega ^{\prime
})g_{n}(r,r^{\prime })Q_{nlm}(\Omega )
\end{equation*}%
where the four-dimensional harmonics $Q_{nlm}$ obey to
\begin{equation*}
-\Delta _{\Omega }Q_{nlm}=(n^{2}-1)Q_{nlm},
\end{equation*}%
($n=1,2,...$ and $\Delta _{\Omega }$ is the angular part of the Laplace
operator) and which are (e.g., see Refs. \cite{fockX})
\begin{equation*}
Q_{nlm}(\Omega )=q_{lm}^{(n)}Y_{lm}(\theta ,\phi )\Pi _{nl}(\chi ),
\end{equation*}%
where $q_{lm}^{(n)}$ are normalization constants, $Y_{lm}(\theta ,\phi )$
are the usual three-dimensional spherical harmonics, and
\begin{equation*}
\Pi _{nl}(\chi )=\sin ^{l}\chi \frac{d^{l+1}(\cos n\chi )}{d(\cos \chi
)^{l+1}},~(l=0,1,...,n-1).
\end{equation*}

In the presence of the wormhole the Green function $G$ contains $G_{0}$ as
an additive part. The function $G_{0}(x,x^{\prime })$ admits the
decomposition as%
\begin{equation*}
G_{0}(x,x^{\prime })=\sum_{n=1}^{\infty }g_{n}^{0}(r,r^{\prime
})Q_{n}(\chi )=\frac{m^{2}}{4\pi ^{2}}\frac{K_{1}(m|x-x^{\prime
}|)}{m|x-x^{\prime }|}
\end{equation*}%
where%
\begin{equation*}
g_{n}^{0}(r,r^{\prime })=\frac{nm^{2}}{2\pi ^{2}}\frac{K_{n}(mr_{>})}{mr_{>}}%
\frac{I_{n}(mr_{<})}{mr_{<}},
\end{equation*}
\begin{equation*}
Q_{n}(\chi )=\frac{2\pi ^{2}}{n}\sum_{l=0}^{n-1}\sum_{m=-l}^{l}Q_{nlm}^{\ast
\prime }(\Omega ^{\prime })Q_{nlm}(\Omega ),
\end{equation*}%
$\chi $ denotes the angle between $r$ and $r^{\prime }$, and $r_{>}$, $r_{<}$
denote the biggest and smallest value of \thinspace $r$ and $r^{\prime }$
respectively. In the massless case we find%
\begin{equation*}
G_{0}(x,x^{\prime })=\frac{1}{4\pi ^{2}\left( x-x^{\prime }\right) ^{2}}=%
\frac{1}{4\pi ^{2}}\sum_{n=1}^{\infty }\frac{r_{<}^{n-1}}{r_{>}^{n+1}}%
Q_{n}(\chi )
\end{equation*}%
which can be used as the generating function for polynomials $Q_{n}$.

The Green function $G=G_{0}+\delta G$ upon the decomposition gives for the
region $r>a$%
\begin{equation*}
G(x,x^{\prime })=\sum_{n=1}^{\infty }\left( g_{n}^{0}(r,r^{\prime })+A_{n}%
\frac{K_{n}(mr)}{mr}\right) Q_{n}(\chi )
\end{equation*}
and for the region $r<a$ ($\widetilde{r}=a^{2}/r>a$)
\begin{equation*}
G(x,x^{\prime })=\sum_{n=1}^{\infty }Q_{n}(\chi )B_{n}\frac{K_{n}(m%
\widetilde{r})}{m\widetilde{r}}.
\end{equation*}%
Now to get the solution to (\ref{weq}) \ we have to match these solutions at
$r=a$, i.e., to require continuity $\frac{\partial g_{n}}{\partial r}$ and $%
g_{n}$, which gives the solution for the region $r>a$
\begin{equation*}
G(x,x^{\prime })=\sum_{n=1}^{\infty }\left( g_{n}^{0}(r,r^{\prime })-\mu
_{n}(a,r^{\prime })g_{n}^{0}(r,a)\right) Q_{n}(\chi )
\end{equation*}%
and as $r<a$
\begin{equation*}
G(x,x^{\prime })=\sum_{n=1}^{\infty }\nu _{n}(a,r^{\prime })g_{n}^{0}(%
\widetilde{r},a)Q_{n}(\chi ).
\end{equation*}
The coefficients $\mu $ and $\nu $ are given by
\begin{equation*}
\mu _{n}(a,r^{\prime })=\frac{\left( \ln \frac{I_{n}(ma)}{ma}\frac{K_{n}(ma)%
}{ma}\right) ^{\prime }}{2\left( \ln \frac{K_{n}(ma)}{ma}\right) ^{\prime }}%
\frac{\frac{K_{n}(mr^{\prime })}{mr^{\prime
}}}{\frac{K_{n}(ma)}{ma}},
\end{equation*}%
\begin{equation*}
\nu _{n}(a,r^{\prime })=\frac{\left( \ln \frac{K_{n}(ma)}{ma}\right)
^{\prime }-\left( \ln \frac{I_{n}(ma)}{ma}\right) ^{\prime }}{2\left( \ln
\frac{K_{n}(ma)}{ma}\right) ^{\prime }}\frac{\frac{K_{n}(mr^{\prime })}{%
mr^{\prime }}}{\frac{K_{n}(ma)}{ma}},
\end{equation*}%
where $\left( \frac{K_{n}(z)}{z}\right) ^{\prime }=\frac{\partial
}{\partial z}\frac{K_{n}(z)}{z}$. We expect that virtual wormholes
have the Planckian size and therefore $ma=x\rightarrow 0$ which
gives asymptotically $ \frac{I_{n}(x)}{x}\rightarrow
\frac{1}{2n!}\left( \frac{x}{2}\right) ^{n-1}$, and
$\frac{~K_{n}(x)}{x}\rightarrow \frac{(n-1)!}{4}\left(
\frac{2}{x}\right) ^{n+1} $. Therefore, the above coefficients are
\begin{equation*}
\mu _{n}(a,r^{\prime })=\frac{1}{\left( n+1\right)
}\frac{4}{(n-1)!}\left( \frac{ma}{2}\right)
^{n+1}\frac{K_{n}(mr^{\prime })}{mr^{\prime }},
\end{equation*}%
\begin{equation*}
\nu _{n}(a,r^{\prime })=\frac{n}{\left( n+1\right)
}\frac{4}{(n-1)!}\left( \frac{ma}{2}\right)
^{n+1}\frac{K_{n}(mr^{\prime })}{mr^{\prime }},
\end{equation*}%
which gives for the region $r>a$
\begin{equation*}
G(x,x^{\prime })=\sum_{n=1}^{\infty }\left( g_{n}^{0}(r,r^{\prime })-\frac{%
\left( \frac{ma}{2}\right) ^{2n}}{\left( n+1\right) \left( (n-1)!\right) ^{2}%
}\frac{m^{2}}{\pi ^{2}}\frac{K_{n}(mr^{\prime })}{mr^{\prime }}\frac{%
K_{n}(mr)}{mr}\right) Q_{n}(\chi )
\end{equation*}%
and as $r<a$%
\begin{equation*}
G(x,x^{\prime })=\sum_{n=1}^{\infty }\frac{n\left( \frac{ma}{2}\right) ^{2n}%
}{\left( n+1\right) \left( (n-1)!\right) ^{2}}\frac{m^{2}}{\pi ^{2}}\frac{%
K_{n}(mr^{\prime })}{mr^{\prime }}\frac{K_{n}(m\widetilde{r})}{m\widetilde{r}%
}Q_{n}(\chi )
\end{equation*}

In the massless case $m=0$ this solution transforms to (as $r>a$)%
\begin{equation*}
G(x,x^{\prime })=G_{0}+\Delta G=\frac{1}{4\pi ^{2}}\sum_{n=1}^{\infty
}\left( \frac{r_{<}^{n-1}}{r_{>}^{n+1}}-\frac{1}{n+1}\frac{1}{a^{2}}\left(
\frac{a^{2}}{rr^{\prime }}\right) ^{n+1}\right) Q_{n}(\chi ).
\end{equation*}
and as $r<a$%
\begin{equation*}
G(x,x^{\prime })=\Delta G=\frac{1}{4\pi ^{2}}\sum_{n=1}^{\infty }\frac{n}{n+1%
}\frac{1}{a^{2}}\left( \frac{a^{2}}{r^{\prime }\widetilde{r}}\right)
^{n+1}Q_{n}(\chi ).
\end{equation*}%
In what follows we shall use the dilute gas approximation and, therefore, it
is sufficient to retain the lowest (monopole $n=1$) term only. Then we find%
\begin{equation}
\Delta G_{\pm }(x,x^{\prime })=\mp 2\pi ^{2}a^{2}G_{0}(r^{\prime
})G_{0}(r_{\pm })
\end{equation}%
where $r_{\pm }>a$ and the sign $\pm $ corresponds to the regions $r_{+}=r>a$
and $r_{-}=a^{2}/r$ ($r<a$) respectively.

\section{Green function in a gas of virtual wormholes}

Consider now the Green function in the presence of a gas of
wormholes. In the previous section we have shown that in the
presence of a single wormhole which connects two Euclidean spaces
the source $\delta \left( x-x^{\prime }\right) $ generates a set
of multipoles placed in the center of sphere (i.e., on the
throat). In the present paper we shall consider a dilute gas
approximation and, therefore, it is sufficient to retain the
lowest (monopole $n=1$) term only. A single wormhole which
connects two regions in the same space is a couple of conjugated
spheres $S_{\pm }^{3}$ of the radius $a$ with a distance
$\vec{X}=\vec{R}_{+}-\vec{R}_{-}$ between centers
of spheres. So the parameters of the wormhole are\footnote{%
The additional parameter (rotation matrix $\Lambda $) is important only for
multipoles of higher orders.} $\xi =(a,R_{+},R_{-})$. The interior of the
spheres is removed and surfaces are glued together. Then the proper boundary
conditions (the actual topology) can be accounted for by adding the bias of
the source
\begin{equation}
\delta (x-x^{\prime })\rightarrow N\left( x,x^{\prime }\right) =\delta
(x-x^{\prime })~+b\left( x,x^{\prime }\right) .
\end{equation}%
In the approximation $a/X\ll 1$ (e.g., see also \cite{KS07X}) the bias takes
the form
\begin{equation}
b_{0}\left( x,x^{\prime },\xi \right) =2\pi ^{2}a^{2}\left( G_{0}\left(
R_{-}-x^{\prime }\right) -G_{0}\left( R_{+}-x^{\prime }\right) \right) \times
\label{X_b1}
\end{equation}%
\begin{equation*}
\times \left[ \delta ^{4}(x-R_{+})-\delta ^{4}(x-R_{-})\right]
\end{equation*}%
We expect that virtual wormholes have throats $a\sim \ell _{pl}$ of the
Planckian size, while in the present paper we are interested in much larger
scales. Therefore, the form (\ref{X_b1}) is sufficient for our aims. However
this form is not acceptable in considering the short-wave behavior and
vacuum polarization effects (e.g., the stress energy tensor). In the last
case one should account for the finite value of the throat size and replace
in (\ref{X_b1}) the point-like source with the surface density (induced on
the throat) e.g., see for details \cite{KS10X}, $\delta ^{4}(x-R_{\pm
})\rightarrow \frac{1}{2\pi ^{2}a^{3}}\delta (\left\vert x-R_{\pm
}\right\vert -a).$

In the rarefied gas approximation the bias function for the gas of wormholes
is additive, i.e.,
\begin{equation}
b_{total}\left( x,x^{\prime }\right) =\sum b_{0}\left( x,x^{\prime },\xi
_{i}\right) =\int b_{0}(x,x^{\prime },\xi )F(\xi )d\xi ,  \label{X_b2}
\end{equation}%
where $F( \xi )$ is the density of virtual wormholes in the
configuration space (i.e., in the space of wormhole parameters)
which is given by
\begin{equation}
F\left( \xi \right) =\sum\limits_{i=1}^{N}\delta \left( \xi -\xi _{i}\right)
.  \label{X_F}
\end{equation}

In the vacuum case we may expect a homogeneous distribution $\rho (\xi )=$ $%
<0|F\left( \xi \right) |0>=\rho \left( a,X\right) $ \footnote{%
Such a function admits a rigorous definition and we consider it in a
separate paper.}, then for the mean bias we find
\begin{equation}
\overline{b}_{total}\left( x-x^{\prime }\right) =\int 4\pi ^{2}a^{2}\left(
G_{0}\left( R_{-}\right) -G_{0}\left( R_{+}\right) \right) \delta
^{4}(x-x^{\prime }-R_{+})\rho \left( a,X\right) d\xi  \label{X_bx}
\end{equation}%
Consider the Fourier transform $\rho \left( a,X\right) =\int \rho \left(
a,k\right) e^{-ikX}\frac{d^{4}k}{\left( 2\pi \right) ^{4}}$ then we find for
$b\left( k\right) =\int b\left( x\right) e^{ikx}d^{4}x$ the expression
\begin{equation}
\overline{b}_{total}\left( k\right) =\frac{4\pi ^{2}}{k^{2}+m^{2}}\int
a^{2}\left( \rho \left( a,k\right) -\rho \left( a,0\right) \right) da,
\label{X_bk}
\end{equation}%
which forms the background cutoff function $\overline{N}\left( k\right) =1+%
\overline{b}_{total}\left( k\right) $, so that the regularized vacuum Green
function $G_{reg}\left( k\right) $ has the form%
\begin{equation}
G_{reg}\left( k\right) =\frac{1}{k^{2}+m^{2}}\overline{N}\left( k\right) .
\label{X_GF}
\end{equation}%
General properties of the cutoff is that $\overline{N}\left( k\right)
\rightarrow 0$ as $k\gg k_{pl}$. We point out that in the case of a dense
gas \cite{S12ef} one has to account for the multiple scattering $\overline{N}%
\left( k\right) =1+\overline{b}\left( k\right) +\overline{b}^{2}\left(
k\right) +...$ (and in general one has also to add the contribution from
higher order multipoles) which gives $\overline{N}\left( k\right) =1/(1-%
\overline{b}\left( k\right) )$. Then in the the low energy limit $k\ll
k_{pl} $ one finds $\rho \left( a,k\right) \approx \rho \left( a,0\right) +%
\frac{1}{2}\rho ^{\prime \prime }\left( a,0\right) k^{2}$ and $G_{reg}\left(
k\right) $ reduces to the renormalization of coupling constants
\begin{equation*}
G_{reg}\left( k\right) =\frac{Z}{k^{2}+m^{2}Z}
\end{equation*}%
where $Z=1/\left( 1+\beta \right) $ and $\beta =-2\pi ^{2}\int
a^{2}\rho ^{\prime \prime }\left( a,0\right) da$ ($\beta>0$, e.g.,
see Refs. \cite{KS07X,KS10X}) which is in agreement with the
statement in Refs. \cite{Lambda}.

\section{Coherent set of virtual wormholes}

As it was demonstrated in Ref. \cite{S13} a coherent set of virtual
wormholes (e.g., beads of virtual wormholes) may work as an actual wormhole.
In the present section we clarify this relation and give simplest examples.
Consider first the distribution function which corresponds to a particular
wormhole. Such a function is given simply by%
\begin{equation*}
F\left( \xi \right) =\delta \left( a-a_{0}\right) \delta ^{4}\left(
R_{+}-y\right) \delta ^{4}\left( R_{-}-y^{\prime }\right)
\end{equation*}%
which corresponds to the single wormhole with $R_{+}=y$, $R_{-}=y^{\prime }$, and $%
a=a_{0}$. Let us fix the vector $X=y-y^{\prime }=const$ and add
the obvious symmetry for the replacement $R_{+}\longleftrightarrow
R_{-}$ which gives
\begin{equation*}
F\left( \xi ,y\right) =\delta \left( a-a_{0}\right) \frac{1}{2}\left( \delta
^{4}\left( R_{+}-y\right) \delta ^{4}\left( R_{-}-y+X\right) +\delta
^{4}\left( R_{-}-y\right) \delta ^{4}\left( R_{+}-y+X\right) \right) .
\end{equation*}%
This allows us to define coherent sets of wormholes by integrating the above
distribution over some portion of space with a density $n(y)$ as%
\begin{equation}
F\left( \xi ,n\right) =\int F\left( \xi ,y\right) n(y)d^{4}y.  \label{fn}
\end{equation}%
In the limit $a_{0}\rightarrow 0$ the virtual wormhole degenerates into a
point and the above expression for $n(y)=\delta \left( f(y)\right) $ defines
merely the gluing of the hypersurface \thinspace $S=$ $\{f(y)=0\}$ and the
shifted surface $S^{\prime }$ by the rule that every point on $\,S$, i.e., $%
y\in S$ is glued to a point on \thinspace $S^{\prime }$, i.e., to $y^{\prime
}=y-X\in S^{\prime }$. In a more general case when $a_{0}=const\neq 0$ the
density $n(y)<1$ (we have the obvious restriction that every wormhole cuts
the volume $2\pi ^{2}a_{0}^{4}$ in space) and the above density corresponds
to a coherent set of wormholes which glue roughly the domains $D=\{n(y)\neq
0\}$ and the analogous domain $D^{\prime }$ shifted in space on the vector $X
$. The density $n(y)$ here defines the degree of gluing (the transparency
coefficient for the resulting wormhole).

Considering now different densities $n(y)$ we may get different wormholes.
It is supposed that an actual wormhole exists for all times and therefore
the density $n(y)$ does not depend on the time-like part of $y=(t,\vec{r})$
(or $n\neq 0$ for a sufficiently big interval of time $t$). The simplest
example is the analog of an \textit{astrophysical wormhole} which can be
modelled by a spherically symmetric function e.g., $n(y)=n(\vec{r})=n\theta
(r-b)$, where $b$ is related to the radius of the throat and $r=\left\vert
\vec{r}\right\vert $. Another simple example can be called the \textit{%
Star-Gate }when the density $n(\vec{r})$ is concentrated on a
two-dimensional disk, e.g., $n(\vec{r})=n(\rho ,\varphi ,z)=n\theta (\rho
-b)\delta (z)$ (where $\rho ,\varphi ,z$ are polar coordinates of the vector
$\vec{r}$). In the homogeneous case $n(y)=n$ and we get the expression below
(\ref{X_NF}).

Substituting different $n(y)$ into (\ref{fn}) and (\ref{fn}) into (\ref{X_b2}%
) we may define the resulting Green function and define the propagation of
particles in the space with such a gluing.

\section{Additional distribution of wormholes and anisotropy in the speed of
light}

\ Here we consider a more simple situation when additional wormholes are
incoherent but simply have an anisotropic distribution. Let we get a
perturbation in the background (vacuum) distribution of virtual wormholes
with a particular distribution function $f\left( a,X\right) =\delta \rho
(\xi )$ of the form
\begin{equation}
f\left( a,X\right) =\delta n\delta \left( a-a_{0}\right) \frac{1}{2}\left(
\delta ^{4}\left( X-r_{0}\right) +\delta ^{4}\left( X+r_{0}\right) \right) ,
\label{X_NF}
\end{equation}%
where $\delta n=\delta N/V$ is the change in the density of wormholes. We
point out that in the vacuum case the background density of wormholes is
always positive $n\geq 0$, while the value $\delta n$ admits both signs. The
above distribution corresponds to a set of wormholes with the throat $a_{0}$%
, oriented along the same direction $r_{0}$ and with the distance between
throats $r_{0}=\left\vert R_{+}-R_{-}\right\vert $. Then $f\left( a,k\right)
$ reduces to
\begin{equation*}
f\left( a,k\right) =\delta n\delta \left( a-a_{0}\right) \cos \left(
kr_{0}\right) ,
\end{equation*}%
where $\left( kr_{0}\right) =k_{\mu }r_{0}^{\mu }$. Thus from (\ref{X_bk})
we find
\begin{equation}
\delta b\left( k\right) =-\delta na^{2}\frac{4\pi ^{2}}{k^{2}+m^{2}}\left(
1-\cos \left( kr_{0}\right) \right) .  \label{X_B}
\end{equation}

Consider the structure of the bias of the unit source (\ref{X_B}) in the
coordinate representation. Substituting (\ref{X_NF}) into (\ref{X_bx}) we
find%
\begin{equation}
\delta b\left( x\right) =2\pi ^{2}\delta na^{2}\left(
G_{0}(x+r_{0})+G_{0}(x-r_{0})-2G_{0}(x)\right) .
\end{equation}%
We recall that here $G_{0}\left( x-x^{\prime }\right) $ is the
standard Euclidean Green function which, upon the continuation to
the Minkowsky space, transforms to the Feynman propagator which is
important in quantum field theory. However when considering the
propagation of signals we should use the retarding Green function,
e.g., in the massless case
\begin{equation*}
G_{0}\ \rightarrow \ G_{ret}\left( x,x^{\prime }\right) =\frac{1}{R}\delta
(t^{\prime }-t+\frac{1}{c}R),\
\end{equation*}%
while the bias has the same structure (e.g., see \cite{KSS09X} ). Thus we
see that the additional source represents three outgoing spherical waves
which originate at positions $x=0$ and $x=\pm r_{0}$. If $r_{0}$ has only
spatial direction, the additional source $b\left( x\right) $ will form the
wavefront which may overrun the standard wave in the direction $r_{0}$ which
may lead to an anomaly shift $\Delta t=r_{0}/c$ in the duration of the
propagation of signals. The intensity of such an additional signal is
described by the portion of the primary signal scattered on additional
virtual wormholes which is given by $b=|\int \delta b\left( x\right) d^{4}x|$%
\begin{equation}
b=2\pi ^{2}|\delta n|a^{2}r_{0}^{2}\leq 1.  \label{X_bamp}
\end{equation}%
We recall that the dilute gas approximation requires $2\pi ^{2}|\delta
n|a^{4}\ll 1$, i.e., the portion of the volume cut by virtual wormholes
should be sufficiently small, while the ratio $r_{0}/a$ may be an arbitrary
parameter and therefore in general $b$ may reach the order of unity.

The above structure of the additional source does not guarantee
that actual signals may overrun the basic signal \footnote{I am
obligated to V.A. Berezin for pointing out to this fact}. In the
first place it shows that the phase velocity may exceed the speed
of light. The physical meaning however has the group velocity only
which can be found from the dispersion
relations, i.e., from poles of the Green function%
\begin{equation*}
\frac{1}{G\left( k\right) }\approx \left( k^{2}+m^{2}\right)
(1-b(k))=(k^{2}+4\pi ^{2}\delta na^{2}\left( 1-\cos \left( kr_{0}\right)
\right) +m^{2})=0
\end{equation*}%
which in the long-wave or low energy limit ($kr_{0}\ll 1$) becomes
\begin{equation*}
\frac{1}{G\left( k\right) }\approx k_{\perp }^{2}+(1+2\pi ^{2}\delta
na^{2}r_{0}^{2})k_{\parallel }^{2}+m^{2}=0
\end{equation*}%
where $k_{\perp }$ and $k_{\parallel }$ denote orthogonal and longitudinal
components of the wave vector with respect to the direction $r_{0}$. We
recall that in the Minkowsky space one should replace $k_{0}\rightarrow
i\omega $, so the poles become real.

Let $r_{0}$ has only spatial direction $r_{0}=(0,\vec{r}_{0})$. Then we see
that the above dispersion relation corresponds to a particle with an
anisotropic speed of light
\begin{equation*}
c_{\perp }=1,\ \ c_{\parallel }^{2}=1+2\pi ^{2}\delta na^{2}r_{0}^{2}.
\end{equation*}%
Whether the longitudinal light velocity $c_{\parallel }$ exceeds or not the
vacuum value ($c_{\parallel }>1$ or $c_{\parallel }<1$) depends on the sign
of the perturbation in the vacuum density of wormholes $\delta n$. If $%
\delta n>0$, then $c_{\parallel }>1$ and the speed of light will exceed the
vacuum value, i.e., such a medium (gas of virtual wormholes with the
perturbation (\ref{X_NF})) will possess an anomalous dispersion. We point
out that in this case we cannot speak on a superluminality (or
subluminality), e.g., as it is discussed in Ref. \cite{R13}, since the above
change holds for all particles and it merely changes the causal structure of
space.

Consider the case when $r_{0}$ has the time-like direction. Then the speed
of light\ remains isotropic but changes according to
\begin{equation}
c^{2}=1/(1+2\pi ^{2}\delta na^{2}r_{0}^{2})\simeq 1-2\pi ^{2}\delta
na^{2}r_{0}^{2}.  \label{sp}
\end{equation}%
Here we get the inverse situation, the speed of light increases $c>1$, if $%
\delta n<0$, and it decreases $c<1$ when $\delta n>0$. In this
case masses of particles also renormalize as $m^{2}\rightarrow
m^{2}(1-2\pi ^{2}\delta na^{2}r_{0}^{2})$.

The interpretation of the physical mechanism which cause the renormalization
in the speed of light is rather simple. Indeed let $\delta n>0$, then the
spatial vector $r_{0}=(0,\vec{r}_{0})$ defines an instant transport of a
portion of the basic signal (via virtual wormholes) in the direction $\vec{r}%
_{0}$. Due to homogeneous distribution of virtual wormholes (\ref{X_NF})
such a mechanism works everywhere in space and this leads to the
renormalization of longitudinal speed of light which becomes $c_{\parallel
}>1$. On the contrary a time-like vector $r_{0}=(T,0)$ defines a retarding $%
T $ (by wormholes) in the propagation of signals which leads to the common
decrease of the speed of light $c<1$. What kind of situation is realized in
the presence of external classical fields $\varphi _{ext}$ depends on the
exact relation between $\delta n$ and $\varphi _{ext}$.

In conclusion we point out that the external field may have a
complex cosmological evolution which gives quite naturally rise to
a non-trivial variation of interaction constants \cite{K05} (for
discussions of the variation of the fine structure constant see
also Refs. \cite{alp}) and, in particular, to the variable speed
of light cosmology \cite{cvar}.

I acknowledge V.A. Berezin and A.A. Kirillov for useful comments
and discussions.

\end{document}